\documentclass{article}
\usepackage{smc}
\usepackage{times}
\usepackage{ifpdf}
\usepackage[english]{babel}
\usepackage{cite}

\usepackage{amsmath,url,svg}

\def\papertitle{CREPE Notes: A new method for segmenting pitch contours into discrete notes}
\def\firstauthor{Xavier Riley}
\def\secondauthor{Simon Dixon}

\newif\ifpdf
\ifx\pdfoutput\relax
\else
   \ifcase\pdfoutput
      \pdffalse
   \else
      \pdftrue
\fi

\ifpdf 
  \usepackage[pdftex,
    pdftitle={\papertitle},
    pdfauthor={\firstauthor, \secondauthor},
    bookmarksnumbered, 
    pdfstartview=XYZ 
   ]{hyperref}

  \usepackage[figure,table]{hypcap}

\else 
  \usepackage[dvips,
    bookmarksnumbered, 
    pdfstartview=XYZ 
  ]{hyperref}  

  \usepackage[dvips]{epsfig,graphicx}
  \graphicspath{{./figures/}}
  \DeclareGraphicsExtensions{.eps}

  \usepackage[figure,table]{hypcap}
\fi

\hypersetup{
    colorlinks,%
    citecolor=black,%
    filecolor=black,%
    linkcolor=black,%
    urlcolor=black
}

\title{\papertitle}

\twoauthors
  {\firstauthor} {C4DM, Queen Mary University of London \\ %
    {\tt \href{mailto:j.x.riley@qmul.ac.uk}{j.x.riley@qmul.ac.uk}}}
  {\secondauthor} {C4DM, Queen Mary University of London \\ %
    {\tt \href{mailto:s.e.dixon@qmul.ac.uk}{s.e.dixon@qmul.ac.uk}}}

\begin{document}
\capstartfalse
\maketitle
\capstarttrue
\begin{abstract}
Tracking the fundamental frequency (f0) of a monophonic instrumental performance is effectively a solved problem with several solutions achieving 99\% accuracy. However, the related task of automatic music transcription requires a further processing step to segment an f0 contour into discrete notes. This sub-task of note segmentation is necessary to enable a range of applications including musicological analysis and symbolic music generation. Building on CREPE, a state-of-the-art monophonic pitch tracking solution based on a simple neural network, we propose a simple and effective method for post-processing CREPE's output to achieve monophonic note segmentation. The proposed method demonstrates state-of-the-art results on two challenging datasets of monophonic instrumental music. Our approach also gives a 97\% reduction in the total number of parameters used when compared with other deep learning based methods.
\end{abstract}

\section{Introduction}\label{sec:introduction}
\label{sec:intro}

Music can be represented on a number of levels; these range from low level representations, such as an audio waveform, through to various high level features such as pitch
or amplitude, enabling discrete representations of pitch such
as MIDI or sheet music. Moving between these domains successfully is one of the key challenges of Automatic Music Transcription (AMT)~\cite{benetos_automatic_2019}. AMT is a foundational
task for working with music collections, especially when considering the
unprecedented scale of such collections today. These collections otherwise require manual annotation from experts, which is a limiting factor in how they can be used. 

Many good solutions for extracting high level features have been proposed in the literature: for the widely studied task of monophonic f0 tracking a number of methods score above 99\% \cite{crepe,pyin} while the task of onset detection has similarly effective methods available \cite{onsets}. Our method builds on CREPE~\cite{crepe}, which represents the state-of-the-art for monophonic pitch tracking. CREPE uses a small convolutional neural network to process audio samples and output an f0 estimate in a framewise fashion.

The next step to combine these feature-level representations into a discrete representation of pitch is more challenging. Taking an extracted f0 contour and segmenting it using the output of an onset detection algorithm provides a good baseline approach~\cite{faghih}, however there are cases where typical percussive onsets are absent at the start of a note transition, for example in legato passages~\cite{collins2005using}. Other methods make use of vocal specific features such as phoneme boundaries~\cite{yukun} or statistical models applied to pitch contours, as reviewed in ~\cite{flamenco}.

For this work we examine the problem of transcribing monophonic instrumental music into discrete notes with high accuracy. Monophonic note transcription has tended to focus on the human voice as input~\cite{pyin, tony, faghih, flamenco} however the human voice is comparatively limited in terms of range and speed of note transitions when compared with instrumental music, which we consider to be a more challenging target. Early work on this task by Brossier~\cite{brossier2004fast} and Collins~\cite{collins2005using} identifies an important observation: a change in pitch by more than a semitone often demarcates note boundaries. This idea is explored more recently by Faghih et al.~\cite{faghih}. Our method also uses this principle, however it differs in that we combine additional information with the pitch contour to increase the accuracy of the output.

\section{Method}
\label{sec:method}

\begin{figure*}
    \centering
    \fbox{\includegraphics[scale=0.605]{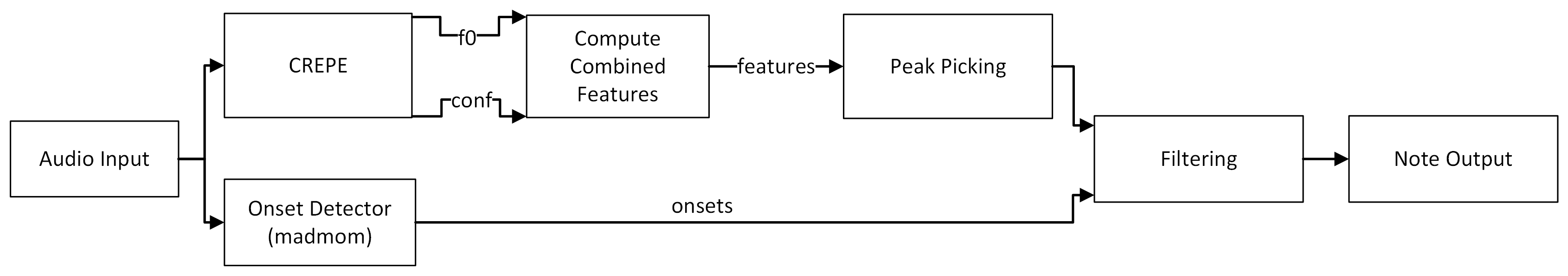}}
    \caption{Flow diagram describing the main stages of the proposed method.}
    \label{fig:flow_diagram}
\end{figure*}

We have named the proposed method CREPE Notes, as it builds on the f0 tracking system CREPE~\cite{crepe}. The steps that make up the method are described in detail below.  Fig.~\ref{fig:flow_diagram} is also included as a visual guide to these steps.

\subsection{The CREPE model}

CREPE is a state-of-the-art solution for monophonic pitch tracking. It processes raw audio samples using a convolutional neural network (CNN) to predict an f0 value for every 10ms of a given piece of audio. F0 predictions are made for 360 bins, each representing a 20 cent step across 6 octaves, and are trained using a binary cross entropy loss.

In addition to f0 estimates, CREPE also includes a measure of ``confidence" which estimates the strength of the pitch content relative to the overall signal. This confidence measure is one of the inputs to our method. We observe that there are typically troughs of varying magnitude at note transitions, however the confidence signal is noisy and simple thresholding is not usually effective. E.g., see Fig.~\ref{fig:data}(b), where the troughs (marked with circles) that correspond to note boundaries in Fig.~\ref{fig:data}(a) have a wide range of confidence values, overlapping with non-boundary trough values.

\subsection{Pitch gradient}

As previously discussed, the rate of change in the f0 contour can also be used as an indicator of note boundaries~\cite{brossier2004fast,collins2005using,faghih}. When making use of an f0 contour it is important to note that pitch is perceived in a roughly logarithmic fashion. This means that the f0 gradient for a unit pitch change will differ depending on the pitch: e.g., B0$\rightarrow$C1 $\approx$ 2\,Hz, while B7$\rightarrow$C8 $\approx$ 236\,Hz. Faghih et al.\ address this by ``stretching the pitch contour"~\cite{faghih}; we achieve a similar result by converting the f0 estimates in Hz to semitone units, expressed as a fractional MIDI number, which is based on log frequency.

An example of such a pitch gradient is shown in Fig.~\ref{fig:data}(c) with peaks in the gradient roughly corresponding to note boundaries shown as circles. Similar to the confidence measurements, we note that these peaks are not at consistent heights and it would be difficult to choose a threshold value that distinguishes real from false note boundaries.

\subsection{Initial segmentation}
\label{ssec:segmentation}

\begin{figure}

\begin{minipage}[t]{1.0\linewidth}
  \centering
  \centerline{\includegraphics[width=\textwidth,trim=0 10 0 10,clip]{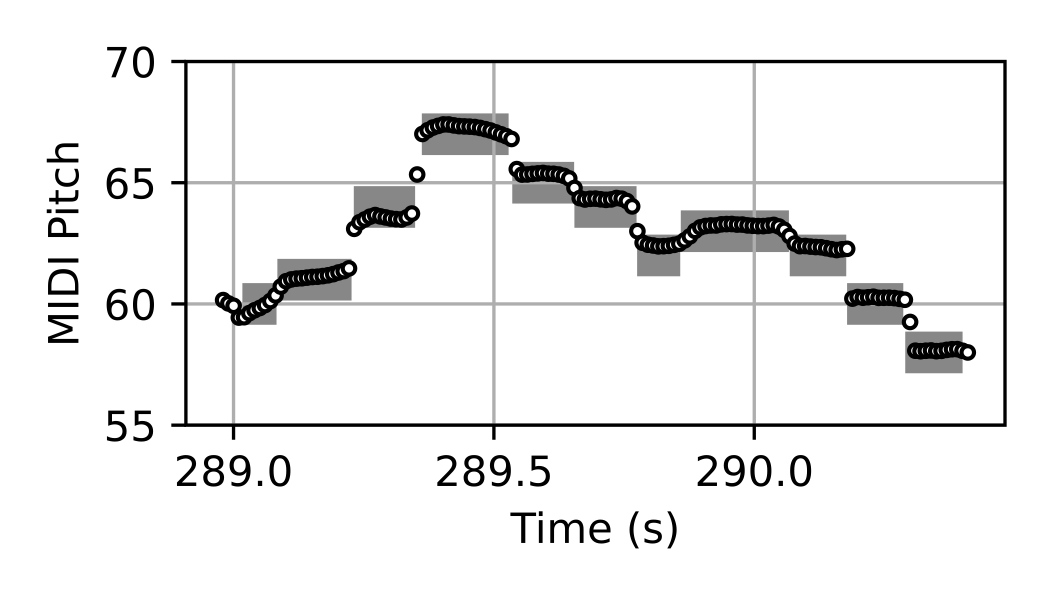}}
  \begin{center}
  (a) Ground truth MIDI note annotations (grey boxes) and CREPE pitch estimates (circles)
  \end{center}
\end{minipage}
\begin{minipage}[t]{1.0\linewidth}
  \centering
  \vspace*{1.7mm}
  \centerline{\includegraphics[width=\textwidth,trim=0 10 0 0,clip]{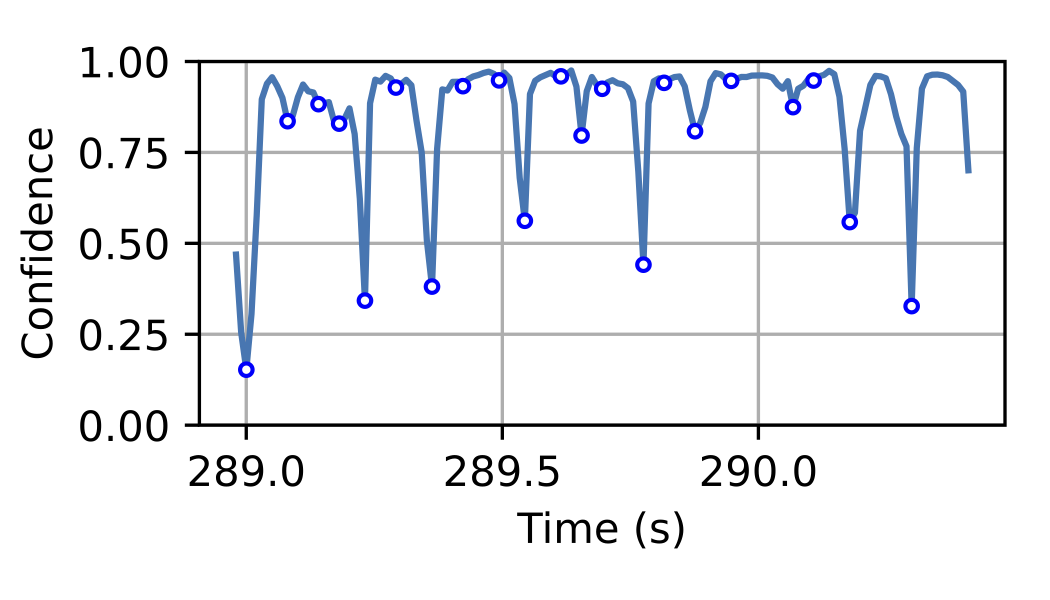}}
  \begin{center}
  (b) CREPE Confidence (blue line) with troughs (circles)
  \end{center}
\end{minipage}
\begin{minipage}[t]{1.0\linewidth}
  \centering
  \vspace*{2.7mm}
  \centerline{\includegraphics[width=\textwidth,trim=0 8 0 0,clip]{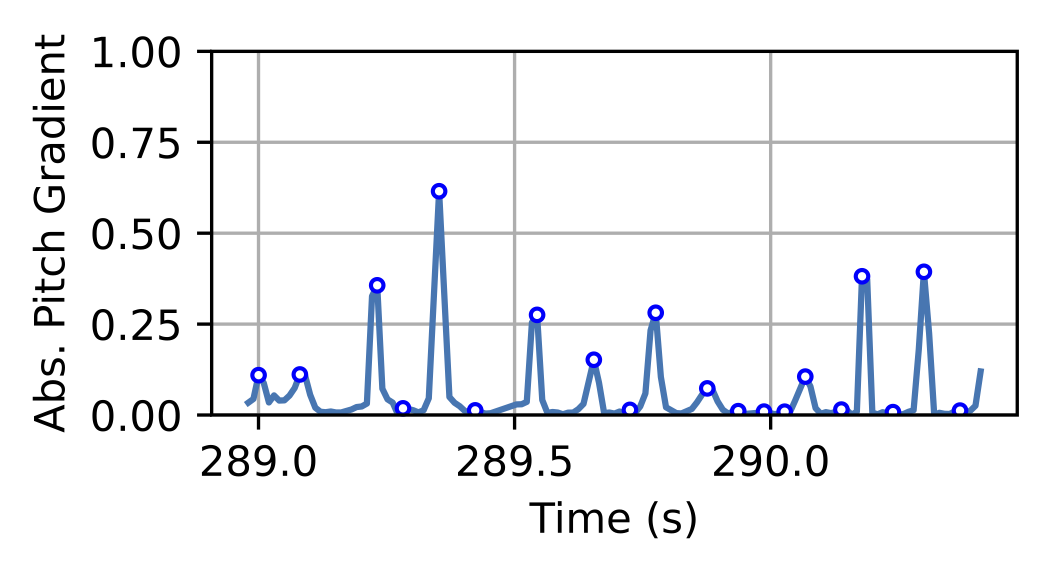}}
  \begin{center}
  (c) Absolute pitch gradient (blue line) with peaks (circles)
  \end{center}
\end{minipage}
\begin{minipage}[t]{1.0\linewidth}
  \centering
  \vspace*{2.7mm}
  \centerline{\includegraphics[width=\textwidth,trim=0 8 0 0,clip]{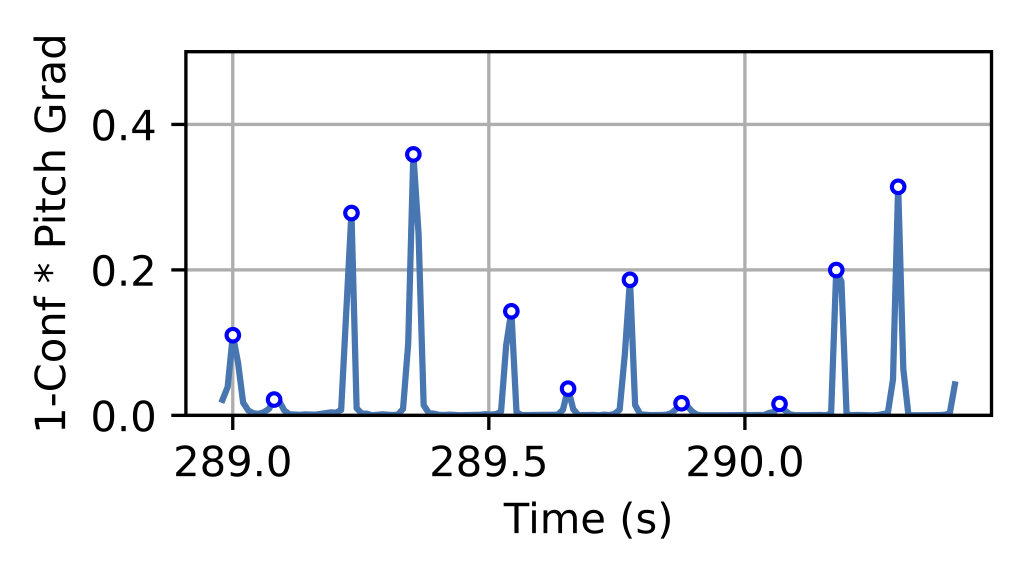}}
  \begin{center}
  (d) Combined confidence and gradient (blue line) with peaks (circles)
  \end{center}
\end{minipage}
\caption{Data and features for an extract from the Filosax dataset (Participant 4, Track 17). X-axis shows time in seconds.}
\label{fig:data}
\end{figure}

Neither the confidence metric from CREPE, nor the peaks in the pitch gradient, are directly usable for segmentation without some variable threshold being chosen. A contribution of this work is that a cleaner signal can be obtained by combining these two features. We take the inverted confidence output from CREPE
(turning valleys into peaks) and multiply it by the normalised absolute gradient of the f0
contour. An example can be seen in Fig.~\ref{fig:data}(d) -- we would like to highlight that the combined signal exhibits less noise and better defined peaks than either of the component signals alone. Given the clarity of this combined signal, a low threshold (empirically set at 0.002 by default) can be used to detect peaks which generally correspond to note boundaries. The detected peaks (shown as circles) align well with the ground truth note segmentations shown in Fig.~\ref{fig:data}(a).

This approach does incur some false negatives, particularly around boundaries for repeated notes, where the pitch gradient is close to zero. These are addressed via an alternative mechanism discussed in Section \ref{ssec:repeated_notes}.

A further processing
step compares each pair of successive segments. If the median pitch for the segments differs by more than 1 semitone then the boundary between the segments is
confirmed. But if the difference in median pitch between the two segments is less than one semitone, then the segments are combined to form a longer segment. This helps to avoid false positives in the segmentation process. The result is a sequence of candidate notes. The use of the median pitch for this step follows the implementation of the widely used PYIN Notes function~\cite{pyin}.

\subsection{Repeated notes}
\label{ssec:repeated_notes}

The steps mentioned above assume that every onset coincides with a pitch transition, but this is not the case for repeated notes. For these we see  no movement in the f0 gradient, and if we see insufficient change in the confidence measure from CREPE, it will result in the segments being merged. For the specific case of segmenting repeated notes we employ an additional onset detection algorithm~\cite{onsets} with a very high threshold (empirically set at 0.7) and re-segment any long notes that contain highly probable onsets within them. Treating repeated notes as a separate step in the transcription process is also described in \cite{kong2021high} which focuses on piano transcription.

\subsection{Amplitude thresholding and trimming}
\label{ssec:amp_thresholding}

At this stage the note segmentation is practically useful, however the original CREPE method does not explicitly distinguish between periods of silence and instrumental activity. This leads to spurious predictions during periods of silence or inactivity. To remove these, we compute the maximum amplitude measurement for each note segment and remove those below a user-configurable threshold from 0 to 127. This corresponds to the velocity parameter used in the MIDI protocol and is set by default to 15. We also remove extremely short notes via a configurable threshold, set by default at 30ms (determined empirically).

A final stage of processing is required because CREPE is insensitive to the amplitude profile of a signal, which means the detection of pitch can occur at very low levels. While the pitch may technically be present, it might not align well with the perceptual onset or offset. We adjust the remaining onsets and offsets in a step we call ``amplitude trimming", where each predicted note has the start and end trimmed if the amplitude has not reached or has fallen below a user-configured threshold.

\section{Experiments}\label{sec:experiments}

\subsection{Datasets}
\label{ssec:datasets}

While the task of f0 tracking can make use of synthetic datasets to obtain accurate ground truth~\cite{crepe}, the process of note annotation on real audio often requires human input. The challenge of synthesising data on a note level is that the variations in timbre and amplitude found in real recordings are difficult to imitate reliably.

Work by Wu et al.\ \cite{mididdsp} offers the possibility of producing such a synthetic dataset with greater levels of realism along with accurate note annotations. This approach has very recently been adopted in the ``Chamber Ensemble Generator" (CEG) dataset \cite{ceg}. The work is promising but CEG focuses on polyphonic examples with relatively slow moving parts (e.g.\ chorales), making it less suited to our task of solo instrumental onset and offset evaluation. Generating more challenging synthetic datasets for solo instrumental material is left for future work.

While several datasets with ground truth note annotations are available, we avoided those that were used to train the two other deep learning based methods (Basic Pitch and MT3 - see below). Instead we test these approaches on two ``unseen'' instrumental datasets to facilitate a fair comparison.

We would like to focus on the performance of the methods under challenging conditions, for example pieces which contain short notes, rapid changes and/or legato phrasing with unclear onsets. The Filosax dataset \cite{filosax} is ideal for this purpose as it contains around 24 hours of audio, made of jazz saxophone solos with expert human note annotations.

We also show results for the same methods on the ``ITM-Flute-99'' dataset: a smaller corpus of 99 recordings of traditional Irish flute \cite{kokuer_curating_2019} totalling around 20 minutes of audio. This dataset is particularly challenging due to the high number of short notes (as short as 30ms) which are integral to the style of the music. The recordings in this dataset did not all tune to a 440\,Hz standard which caused errors in the ground truth pitch labels provided by the authors. To rectify this, we converted all recordings to use 440\,Hz tuning using Librosa\footnote{\url{https://librosa.org/}} to estimate the original tuning and RubberBand\footnote{\url{https://breakfastquay.com/rubberband/}} for high quality pitch shifting. Ground truth note annotations were also provided in \,Hz and these were scaled to the new tuning before the corrected pitch labels were re-calculated.

\subsection{Methodology}
\label{ssec:methodology}

We compare CREPE Notes with a number of other systems, however we note that monophonic instrumental note segmentation is addressed relatively infrequently in the existing literature. PYIN~\cite{pyin} initially published results on f0 tracking, however the official implementation also contains code to segment a signal into notes which has been widely cited and adopted in dataset creation~\cite{bittner_vocadito_nodate, xi_guitarset_2018}.
Bittner et al.~\cite{Bittner} have recently proposed an instrument agnostic, multipitch tracking method which also works for monophonic transcription. We make use of their latest release of this work which is packaged as ``Basic Pitch".\footnote{\url{https://basicpitch.spotify.com/}}

We also include results for MT3, proposed by Hawthorne et al.~\cite{mt3}. This is a system designed for multi-instrument, polyphonic transcription that also functions for the monophonic instrumental case. The results are obtained using the published model, however it was not trained on saxophone or flute data specifically so it is likely that results could be improved with a more specific model or fine tuning.

We were unable to compare with the method proposed by Faghih et al.~\cite{faghih} directly, as they focus on vocal transcription and we were unable to run their published implementation. A full comparison is a goal for future work.


A range of model sizes are available for CREPE, and a parameter count for the published models is listed where applicable. While the proposed method does not use deep learning directly, we quote the size of the CREPE model which was used to provide the f0 and confidence estimates for these experiments.  We include results for the largest and smallest available models, named ``full'' and ``tiny'' respectively.

\subsection{Results}
\label{ssec:results}

We use the \texttt{mir\_eval}~\cite{mireval} library to calculate precision, recall, F-measure and overlap. A default threshold of 50ms was used to evaluate onset and offset accuracy and f0 accuracy is tracked implicitly as f0 errors will result in lower scores. Results across 238 tracks in Filosax\footnote{Two tracks omitted due to alignment issues in ground truth annotations.} are shown in Table~\ref{table:results}. Results for the ``ITM-Flute-99'' dataset are shown in Table~\ref{table:itm-results}.

\begin{table}
\begin{tabular}{r|l|l|l|l|l}
 & CNt             & CN             & PYIN  & BP    & MT3   \\
\hline
Recall               & 88.26           & \textbf{88.61} & 50.32 & 80.62 & 40.67 \\
Precision            & \textbf{77.18}  & 76.91          & 69.50 & 71.18 & 45.78 \\
F-measure            & \textbf{82.31}  & \textbf{82.31} & 58.28 & 75.54 & 42.97 \\
Overlap              & 88.54           & \textbf{89.91} & 87.36 & 83.45 & 72.96 \\
\hline
Parameters           & 0.5M            & 22M            & N/A   & 17K   & 77M

\end{tabular}

\caption{\label{table:results}%
Results on the Filosax dataset.
Mean scores are shown for each metric. Abbreviations are CNt (Crepe Notes ``tiny" model, proposed), CN (Crepe Notes ``full" model, proposed), PYIN (PYIN Notes), BP (Basic Pitch). Parameter counts for each model are shown for reference. For the proposed models we quote the size of the CREPE model which was used to provide the f0 and confidence estimates.}
\label{fig:res}

\end{table}

\begin{table}
\begin{tabular}{r|l|l|l|l|l}
 & CNt             & CN             & PYIN\footnotemark  & BP    & MT3   \\
\hline
Recall               & \textbf{66.66}  & 65.79 & 36.58 & 55.56 & 23.87 \\
Precision            & 66.73  & \textbf{67.18} & 64.83 & 64.92 & 28.35 \\
F-measure            & \textbf{66.58}  & 66.35 & 46.44 & 59.58 & 25.47 \\
Overlap              & 79.96  & 80.53 & \textbf{82.50} & 77.33 & 69.02 \\
\hline
Parameters           & 0.5M            & 22M            & N/A   & 17K   & 77M

\end{tabular}

\caption{\label{table:itm-results}%
Results on the ITM Flute 99 dataset, showing mean scores for each metric. Abbreviations are given in Table~\ref{table:results}.}
\label{fig:itm-results}

\vspace{-4mm}

\end{table}

\footnotetext{Dataset was originally created with PYIN, thus the high overlap score.}

The proposed method outperforms the others we have examined for these datasets. From informal observations, all methods appear to produce good results on slower moving passages but the proposed method maintains a high degree of accuracy when faster groups of notes are played, making it suitable for solo instrumental transcription tasks.

Surprisingly, the use of the smallest CREPE model has minimal impact on performance for this task. This can be explained by two factors: one is that the CREPE authors note that there is around a 1.5\% drop in raw pitch accuracy (RPA) between the largest and smallest models which is not a large reduction relative to the difference in parameter count. Also we suspect that taking the median f0 estimate for each segment makes the method more robust to short-term f0 errors.





\section{Discussion and Conclusions}
\label{sec:conclusions}

We present a system for monophonic musical note transcription which outperforms other solutions by a significant margin on two datasets of real instrumental audio. An appealing aspect of the solution is that it is implemented as a series of simple algorithmic post-processing steps over the output of a state-of-the-art f0 tracking system. Once the initial pitch tracking has taken place, the note segmentation runs faster than realtime on a single CPU with no additional models required. The original CREPE model is instrument agnostic~\cite{crepe} and effective on the full range of musical pitch; these are properties which our method inherits. CREPE also reports good robustness to noise which is a property we hope to explore in further work.

The fact that our proposed method is able to outperform a larger model like MT3 may be surprising. However, we would like to highlight that MT3 performs both polyphonic transcription and multi-instrument classification. These tasks together are inherently more difficult and explain the need for the additional parameters. Thereforre MT3 may still be considered to be efficient in that context.


This method currently relies on the use of CREPE as it provides a high level of f0 accuracy and useful signal for segmentation in the form of the confidence measure. Other f0 tracking methods have metrics that are similar to this confidence measure, such as voiced/unvoiced predictions. In future work we seek to examine whether this method can be made more general by combining the calculated pitch gradient with the various confidence terms. 


Our results also suggest that the role of post-processing in this task requires further examination. A similar approach is used in many successful transcription models, namely that pitch contours are first calculated in a frame-wise manner and then post-processed to segment the notes. In future work we hope to examine the role of post-processing for transcription tasks in more detail, including an exploration of methods to learn note segmentation as an end-to-end process. MT3 is an example of such a method, however a more specialised model for monophonic instruments may yield better results. 

\begin{acknowledgments}
The first author is a research student at the UKRI Centre
for Doctoral Training in Artificial Intelligence and Music,
supported by UK Research and Innovation [grant number
EP/S022694/1].
\end{acknowledgments}

\clearpage

\bibliography{smc2023template}

\end{document}